\documentclass[preprint,aps,pre,onecolumn,amsfonts,showpacs]{revtex4} 
\pdfoutput=1
\usepackage{bm}
\usepackage{epsfig,amsopn}
\usepackage{graphicx}
\usepackage{amsmath,amssymb}
\usepackage{natbib}
\usepackage{braket}
\renewcommand{\section}[1]{{\par\it #1.---}}
\def\bG{{\bm{G}}}
\def\bg{{\bm{g}}}
\def\bea{\begin{eqnarray}}
\def\eea{\end{eqnarray}}

\def\nn{\nonumber}
\def\om{\omega}
\def\f{\frac}

\def\bZ{\bm Z}
\begin{document}

\title{Landauer formula for phonon heat conduction: relation between energy transmittance and transmission coefficient}
\author{Suman G. Das and Abhishek Dhar}
\affiliation{Raman Research Institute, Bangalore 560080, India}
\date{\today}

\begin{abstract}
The heat current across a  quantum harmonic system 
connected to reservoirs at different temperatures is given by the
Landauer formula, in terms of an integral over phonon frequencies $\omega$, of the energy transmittance $\mathcal{T}(\omega)$.
There are several different ways to derive this formula, for example  using the Keldysh approach or the Langevin equation approach.  
The  energy transmittance $\mathcal{T}(\omega)$ is usually  expressed 
in  terms of nonequilibrium phonon Green's function and it is expected 
that it is related to the transmission coefficient $\tau(\omega)$ of plane 
waves across the system. In this paper, for a one-dimensional set-up of 
a finite harmonic chain connected to reservoirs which are also  semi-infinite 
harmonic chains, we  present a  simple and direct   
demonstration of the  relation between $\mathcal{T}(\omega)$ and 
$\tau(\omega)$. Our approach is easily extendable to the case where both 
system and reservoirs are in higher dimensions and have arbitrary geometries, 
in which case the meaning of $\tau$ and its relation to $\mathcal{T}$ are
    more non-trivial.
\end{abstract}

\maketitle

\section{Introduction}
The Landauer formula gives an exact expression for the current (energy and/or particle) in ``non-interacting'' quantum systems coupled to reservoirs kept at  different
temperatures ( and/or different chemical potentials ). By ``non-interacting'' one refers to systems described by quadratic Hamiltonians. 
It thus includes harmonic crystals where one considers 
energy transport by phonons, and tight-binding Hamiltonians where  there is transport of both charge and energy by electrons. 
The formula for phonon heat current across a harmonic crystal connected to heat baths at temperatures $T_L,T_R$ is given by
\bea
J=\f{1}{2 \pi} \int_{-\infty}^\infty d \om ~\hbar \om~\mathcal{T} (\om)~[f(\om,T_L)-f(\om,T_R)]~, \label{landform}
\label{current}
\eea
where the quantity $\mathcal{T}(\om)$, which we shall refer to as the energy transmittance, can be expressed in terms of appropriate ``nonequilibrium'' phonon's 
Green's functions and $f(\om,T)=1/(e^{\hbar \om/k_BT}-1)$ is the thermal phonon distribution function. 
This Landauer formula for phononic heat current has been derived rigorously using the quantum Langevin equation approach \cite{dhar06,dhar08} as well as the nonequilibrium Green's function (NEGF) approach \cite{wang06,yamamoto06}.

Landauer's original idea was to think of conductance in terms of transmission or scattering of plane waves and 
for the case where the reservoirs or ``leads'' are one dimensional, it is 
expected  that $\mathcal{T}(\om)$ is related to the transmission coefficient $\tau(\omega)$ of plane waves \cite{imry99}. For the case of electron transmission, 
$\mathcal{T}(\omega)$ can again be expressed in terms of nonequilibrium 
 Green's functions \cite{land_elec} and the relation to the transmission coefficient was directly demonstrated through the work 
of Todorov {\emph et al} using scattering theory \cite{todorov93}.
For the case of phonons  we are not aware  of  an explicit proof of 
this relation and that is the main objective of this paper. Here we consider a general one-dimensional finite harmonic chain coupled to 
reservoirs which are themselves semi-infinite ordered harmonic chains and give 
a  fully quantum-mechanical derivation of the relation between $\mathcal{T}$ and $\tau$  and show how the NEGF current formula can be obtained from the transmission  coefficient. 

In our derivation we first note that in the NEGF approach the energy 
transmission $\mathcal{T}$ is expressed in 
terms of a Green's function. This Green's function can be expressed explicitly 
in terms of a product of $2\times 2$ matrices. 
On the other hand the transmission coefficient can be computed by  
constructing appropriate scattering states and this can be done in 
two ways --- (i) a direct
solution of the discrete wave equation which again gives $\tau$ expressed 
in the form of a product of matrices or (ii) by using the Lippmann-Schwinger 
scattering theory to evolve reservoir normal modes and this  gives $\tau$ directly in terms of the Green's 
function. From the forms of these expressions we directly obtain the required 
relations. 
We note that for the case where the reservoirs are not 
one-dimensional chains, but have arbitrary geometries \cite{kosevich95,mingo03,panzer08}, the NEGF expression for $\mathcal{T}(\omega)$ still has the  same form but it is not clear as to how one should compute $\tau$ and how exactly it is related to $\mathcal{T}$. In this 
case the approach using Lippmann-Schwinger scattering theory can still be used to arrive at the required relation.
A model similar to ours was studied recently by 
Zhang {\emph et al} \cite{zhang11} in the context of interfacial thermal transport in atomic junctions and the
 relation between the  NEGF formula for energy transmittance and the transmission coefficient  was
 established  numerically and also exactly for the special case of a single interface. 

The plan of the paper is as follows. In sec.~\ref{negf} we first define the 
model and state some general results for the heat current given by the formalism of nonequilibrium Green's functions. We  then give an explicit expression for 
the form of the Green's function appearing in the energy transmission formula. 
In sec.~\ref{scatt} we consider the transmission of plane waves across the 
system and, using two different approaches, obtain the form of the 
transmission coefficient. The transmission coefficient can also be expressed in terms of the same nonequilibrium Green's function and using this we write the 
relation between it and $\mathcal{T} (\omega)$. This relation is 
then used in sec.~\ref{current} to derive the Landauer formula for heat current. Finally we discuss our results in 
sec.~\ref{discussion}. 

\section{One-dimensional chain connected to one-dimensional baths}
\label{negf}
\begin{figure}
\includegraphics[scale=0.4]{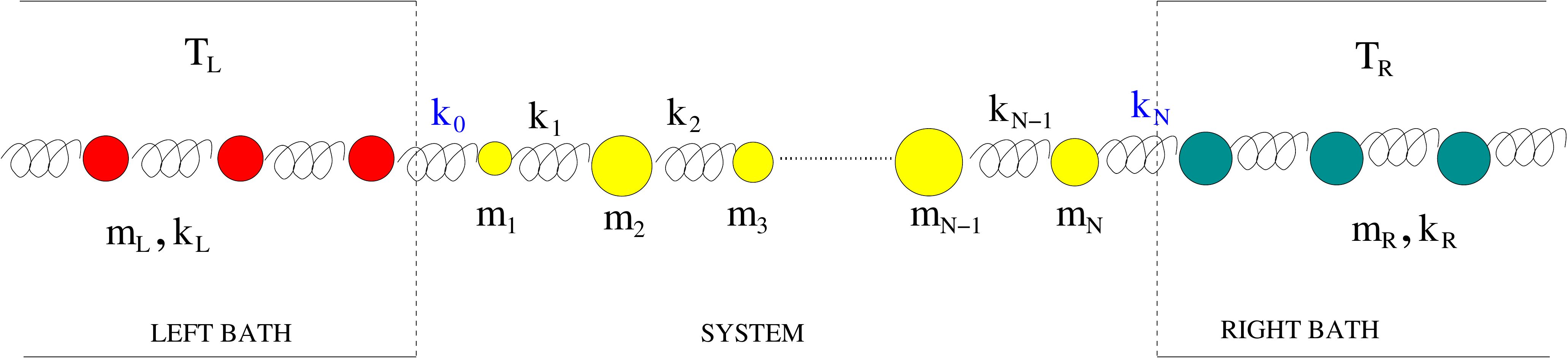}
\caption{Schematic of the set-up considered in the paper. The system consists 
of a harmonic chain of $N$ particles for which both the particle masses and the inter-particle spring constants  
take arbitrary values. The system is sandwiched between two reservoirs which are ordered $1D$ harmonic chains 
with different mass densities and elasticities. The coupling constant between left reservoir and system is $k_0$  
and between right reservoir and system is $k_N$.}
\label{fig1}
\end{figure}
Consider the set-up in Fig.~(\ref{fig1}) where a one-dimenional (1D) harmonic chain with
arbitrary spring constants and masses is connected to leads which are
themselves ordered harmonic chains.  Special cases of this setup have been
discussed earlier by various authors \cite{Rubin71,spohn77,lumpkin78,segal03,dhar12,zhang11} in the 
context of heat conduction. 
Let us assume that the system has $N$
Cartesian positional 
degrees of freedom $\{x_l\}$, $l=1,2\ldots,N$ with corresponding momenta
$\{p_l\}$. These satisfy the usual commutation relations $[x_l,p_m]=i \hbar
\delta_{l,m}$ and $[x_l,x_m]=[p_l,p_m]=0$. Similarly the left reservoir degrees 
of freedom are denoted by
$\{x_\alpha,p_\alpha\}$,~ $\alpha=1,\ldots,N_L$ and the right reservoirs by
$\{x_{\alpha'}, p_{\alpha'}\}$, ~$\alpha'=1,\ldots,N_R$. 
 We consider our system plus reservoir to be described by the full Hamiltonian
\bea
\cal{H} &=& \sum_{l=1}^{N} \frac{p_l^2}{2 m_l} + \sum_{l=1}^{N-1} \frac{k_l
  (x_l-x_{l+1}) ^2}{2}  \nn \\ 
&+& \sum_{\alpha=1}^{N_L} \frac{p_\alpha^2}{2m_L} + \sum_{\alpha=1}^{N_L} \frac{k_L (x_\alpha-x_{\alpha+1})^2}{2} +
\frac{ k_0 (x_{\alpha=1}-x_1)^2}{2} \nn \\
&+&  \sum_{\alpha'=1}^{N_R} \frac{p_{\alpha'}^2}{2m_R} + \sum_{\alpha'=1}^{N_R}\frac{k_R (x_{\alpha'}-x_{\alpha'+1})^2}{2} +
\frac{k_N (x_{\alpha'=1}-x_N)^2}{2} ~, \label{ham1} 
\eea
where we assume $x_{\alpha=N_L+1}=x_{\alpha'=N_R+1}=0$. The system masses
$\{m_l\}$ and
spring constants $\{k_l\}$ are assumed to be arbitrary. The left (right)
reservoir particle masses are all taken to be $m_L$ ($m_R$) and the
inter-particle spring constants are taken to be $k_L$ ($k_R$). To ensure a uniqe steady state We will always assume that the 
reservoirs are chosen to have sufficiently broad bandwidths compared to the
spectrum of the system \cite{dhar06,dharsen06}. 
The above Hamiltonian can
be written in the canonical form:
\bea
\mathcal{H}&=&\mathcal{H}_S+\mathcal{H}_L+\mathcal{H}_R+\mathcal{H}_{LS}+
\mathcal{H}_{RS}~, \label{ham1D}
\eea
where
\bea
\mathcal{H}_S &=&  \sum_{l=1}^{N} \frac{p_l^2}{2 m_l} +
\sum_{l=1}^{N-1} \frac{k_l   (x_l-x_{l+1}) ^2}{2}+\frac{k_0 {x_1}^2}{2}+\frac{k_N {x_N}^2}{2}~,   \nn \\ 
\mathcal{H}_L&=& \sum_{\alpha=1}^{N_L} \frac{p_\alpha^2}{2m_L} + \frac{k_L
  (x_\alpha-x_{\alpha+1})^2}{2} +\frac{k_0 x_{\alpha=1}^2}{2}~, \nn \\
\mathcal{H}_R &=&  \sum_{\alpha'=1}^{N_R} \frac{p_{\alpha'}^2}{2m_R} + \frac{k_R (x_{\alpha'}-x_{\alpha'+1})^2}{2} +
\frac{k_N x_{\alpha'=1}^2}{2} \nn~, \nn \\
\mathcal{H}_{LS}&=& -k_0 x_{\alpha=1} x_1,~~~~~\mathcal{H}_{RS}= -k_N x_{\alpha'=1}x_N ~.
\label{hamcan}
\eea
Using the vector notation $X_S^T=(x_1,x_2,\ldots,x_N)$,~
$P_S^T=(p_1,p_2,\ldots,p_N)$ and similarly $X_L,X_R, P_L,P_R$, the different
parts in the above Hamiltonian can be written as
\bea
\mathcal{H}_S &=& \f{1}{2} P_S^T~{\bm M}_S^{-1} ~P_S +\f{1}{2}
X_S^T ~{\bm K_S}~ X_S~, \nn \\
\mathcal{H}_L &=& \f{1}{2}{P_L}^T~{\bm M}_L^{-1} ~P_L +\f{1}{2}
{X_L}^T ~{\bm K}_L X_L~, \nn \\
\mathcal{H}_R &=& \f{1}{2} {P_R}^T~{\bm M}_R^{-1} ~P_R +\f{1}{2}
{X_R}^T ~{\bm K}_R X_R~, \nn \\
{\mathcal H}_{LS} &=&  X_S^T~{\bm K}_{SL}~X_L~,~~ {\mathcal H}_{RS}=  X_S^T~{\bm  K}
_{SR}~X_R~, \nn
\eea
where ${\bm M}_S,~{\bm M}_L,~{\bm M}_R$ and ${\bm K}_S,~{\bm
  K}_L,~{\bm K}_R$ denote respectively the mass matrix and the
force-constant matrix of the system, left reservoir and right
reservoir, while ${\bm K}_{SL}$ and ${\bm K}_{SR}$ denote the linear
coupling coefficients between the two reservoirs and the system. In our case 
${\bm K}_{SL}$ is a $N \times N_L$ matrix whose only non-zero element is $[{\bm
  K}_{SL}]_{1,1}=k_0$, while ${\bm K}_{SR}$ is a $N \times N_R$ matrix whose
only non-zero element is $[{\bm   K}_{SR}]_{N,1}=k_N$.

{\bf Expression for steady state heat current}: We now consider the situation where at some distant past time ($t< t_0$) the
two reservoirs are uncoupled from the system and are   separately in
equilibrium (and described by canonical distributions) at temperatures $T_L$ and $T_R$ respectively. At time $t_0$ we
start evolving the system plus reservoirs with the full Hamiltonian in
Eq.~(\ref{ham1D}). Eventually we set the reservoir sizes $N_L,N_R \to \infty$
and $t_0 \to -\infty$. The system reaches a nonequilibrium steady
state at finite $t$. Note that  we have included   terms  involving the coupling
coefficients $k_0, k_N$ in the isolated reservoir Hamiltonians. 
As has been discussed using various approaches \cite{dhar06,wang06,yamamoto06}, 
the steady state current can be expressed using the following phonon Green's
function: 
\bea
{\bm G}^{\pm} &=& \frac{1}{-{\bm M}_S \omega^2 + {\bm K}_S - {\bm \Sigma}_L^{\pm} - 
{\bm \Sigma}_R^\pm } \,  ,
\label{GF}
\eea
where the self-energies ${\bm \Sigma}_L^\pm ,{\bm \Sigma}_R^\pm $ can be expressed  in terms of the isolated reservoir 
Green functions ${\bf g}^\pm_{L}(\omega)= [~-{\bm M}_{L}(\omega\pm i
  \epsilon)^2+{\bm 
    K}_L]^{-1}  $ , ${\bf g}^\pm_{R}(\omega)=~[-{\bm M}_{R}(\omega \pm i
  \epsilon)^2+{\bm     K}_R]^{-1}  $
 and the coupling matrices ${\bm K}_{SL},~{\bm K}_{SR} $. The self energies 
are given by 
$ {\bm \Sigma}^\pm_L(\omega) = {\bm K}_{SL}~ {\bf g}^\pm_L(\omega)~ {\bm
    K}_{SL}^{T},~ {\bm \Sigma}^\pm_R(\omega)={\bm K}_{SR}~ {\bm
  g}^\pm_R(\omega)~ {\bm K}_{SR}^{T} ~.$
Defining ${\bm \Gamma}_L(\omega)={\rm Im} [~{\bm \Sigma}^+_L~]~,~{\bm
  \Gamma}_R(\omega)= {\rm Im} [~{\bm \Sigma}^+_R~]$, we find
\cite{dhar06,wang06,yamamoto06} that the steady state current is given by the formula in Eq.~(\ref{landform}) with
\bea
\mathcal{T}(\omega)= 4 Tr[{\bm G}_+(\omega) {\bm
      \Gamma}_L(\omega) {\bm G}_-(\omega) {\bm \Gamma}_R(\omega)]\label{T}~.
\eea
For our one-dimensional system, we note that ${\bm G}^\pm, {\bm \Sigma}^\pm_L,{\bm
  \Sigma}^\pm_R$ are all $N \times N$ matrices. The only non-zero elements of
${\bm \Sigma}^\pm_L$ and ${\bm \Sigma}^\pm_R$ are respectively $[{\bm
    \Sigma}^\pm_L]_{1,1}={k_0}^2 [\bg^\pm_L]_{1,1}=:\Sigma^\pm_L$ and $[{\bm
    \Sigma}^\pm_R]_{N,N}={k_N}^2 [\bg^\pm_R]_{1,1}=:\Sigma^\pm_R$. Let us define
$\Gamma_L=Im[{ \Sigma}^+_L]$,  $\Gamma_R=Im[{ \Sigma}^+_R]$~. 
Hence the expression of $\mathcal{T}$  reduces to:
\bea
\mathcal{T}= 4 \Gamma_L \Gamma_R \bG^+_{1,N}\bG^-_{N,1}=4 \Gamma_L \Gamma_R |\bG^+_{1,N}|^2 ~, \label{Tw}
\eea
with the  matrix ${\bm G}^+=   {\bZ}^{-1} $ , where  $ {\bZ} = -{\bm M}_S \omega^2 + {\bm K}_S~- {\bm \Sigma}_L^+ - {\bm \Sigma}_R^+ $ is 
 a tri-diagonal matrix.
The bandwidth of the two baths are different ($2\sqrt{k_L/m_L}$ and 
$2\sqrt{k_R/m_R }$ for the left and right baths respectively), so the 
conduction of heat across the system will have contribution
only from the overlapping part of the bandwidths. 

{\bf Explicit forms for ${\bm G}^+_{1,N}, \Gamma_L,\Gamma_R $}: Using methods  described in  
\cite{Casher71} we now show that  the Green's function element occuring  in Eq.~(\ref{Tw}) can be expressed in terms of a product of $2 \times 2$ matrices. We
also obtain the explicit forms of $\Gamma_L,\Gamma_R$ for our particular model.
The matrix ${\bm Z}$ has the form
\bea
\bm{Z}(\omega) &=&
 \begin{bmatrix}
   a_1-\Sigma_L^+(\omega)& -k_1 & \cdots  & 0 & 0 & 0\\
   -k_1 & a_2 &-k_2 & \cdots & 0 & 0\\
   \vdots & \vdots &\vdots &\ddots &\vdots &\vdots\\
   0 & 0 & \cdots &-k_{N-2} & a_{N-1} &-k_{N-1}\\
   0 & 0 & 0 & \cdots & -k_{N-1} & a_N-\Sigma^+_R(\omega)\\
                                \end{bmatrix}~, \\
{\rm ~ where~} a_l &=& k_l+k_{l-1}-m_l \omega^2,~~~l=1,\ldots,N~. \nn
\eea
Taking the inverse of this matrix, we get
\bea
{\bm G}^+_{1,N}= \f{\prod\limits_{l=1}^{N-1} k_l}{\Delta_{1,N}}~, \label{G1N}
\eea
where $\Delta_{1,N}$ is defined as the determinant of the matrix ${\bm Z}(\omega)$. Let us also define 
 $ D_{l,j}$  as the determinant of the sub-matrix starting with the $l$-th row and column and ending with the $j$-th row and
column of the matrix $-{\bm M}_S \om^2+{\bm K}_S$.   From the tri-diagonal 
form of the matrices, it is easily shown that
\bea
\Delta_{1,N} &=& (~a_1-\Sigma_L^+~)~[~(a_N-\Sigma_R^+)~D_{2,N-1}-k_{N-1}D_{2,N-2}~] - k_1~[~(a_N-\Sigma_R^+)
D_{3,N-1}-k_{N-1}D_{3,N-2}~]~ \nn \\
 &=& D_{1,N}-\Sigma^+_R D_{1,N-1}- \Sigma^+_L D_{2,N} +\Sigma^+_L \Sigma^+_R D_{2,N-1} \nn \\
&=& \begin{bmatrix}
             1 & -\Sigma^+_L
            \end{bmatrix}
\begin{bmatrix}
             D_{1,N} & -D_{1,N-1}\\
	     D_{2,N} & -D_{2,N-1}\\
            \end{bmatrix}
\begin{bmatrix}
             1\\
             \Sigma^+_R\\
            \end{bmatrix}~.\label{det1N}
\eea
The elements $D_{l,j}$ satisfy the recursion relation $D_{l,N}=a_l D_{l+1,N}-k_l^2 D_{l+2,N}$ for $l=1,...N-2$, and 
$D_{l,N-1}=a_l D_{l+1,N-1} - k_l^2 D_{l+2,N-1}$ for $l=1,...,N-3$. 
In matrix form these give
\bea
\begin{bmatrix}
D_{l,N} & -D_{l,N-1}\\
D_{l+1,N} & -D_{l+1,N-1}\\
\end{bmatrix}
= k_l \begin{bmatrix}
   a_l/k_l & -k_l\\
    1/k_l & 0\\
  \end{bmatrix}
\begin{bmatrix}
D_{l+1,N} & -D_{l+1,N-1}\\
D_{l+2,N} & -D_{l+2,N-1}\\
\end{bmatrix},
\label{recursionD}
\eea
which holds for $l=1,..,N-3$. 
Using these relations and further defining $D_{N+1,N}=D_{N,N-1}=1,~D_{N+2,N}=D_{N+1,N-1}=0$, we arrive at the result 
\bea
\begin{bmatrix}
D_{1,N} & -D_{1,N-1}\\
D_{2,N} & -D_{2,N-1}\\
\end{bmatrix}
&=& \prod_{l=1}^N k_l ~\hat{T}~
\begin{bmatrix}
1 & 0\\
0 & 1/{k_N}^2\\
\end{bmatrix}~, \label{Drec} \\
{\rm where~} \hat{T} \equiv \prod_{l=1}^N \hat{T}_l~,~~~~\hat{T_l}&=&\begin{bmatrix}
a_l/k_l & -{k_l}\\
1/k_l & 0\\
\end{bmatrix}~. 
\label{T_l}
\eea
Hence using Eqs.~(\ref{det1N},\ref{Drec},\ref{T_l}) we get
\bea
\Delta_{1,N}=(\prod\limits_{l=1}^{N} k_l)~ {\begin{bmatrix}
               1 & -\Sigma^+_L\\
              \end{bmatrix}
 ~\hat{T}~\begin{bmatrix}
                                   1 & 0\\
			           0 & 1/{k_N^2}
                                  \end{bmatrix}
                                  \begin{bmatrix}
                                   1\\
			           \Sigma^+_R\\
                                  \end{bmatrix}}~. \label{det1N-2}
\eea
We next  find the explicit forms of $\Sigma^+_L, \Sigma^+_R$, for which we 
need to evaluate the reservoir Green's function elements 
$[{\bf g}_L^+]_{1,1}$ and $[{\bf g}_R^+]_{1,1}$. Consider the left reservoir. 
For the case $k_0=k_L$, it is simple to find all normal modes and hence compute the Green's function corresponding to the 
force matrix ${\bm K}_L={\bm K}_L^0$ (say).  One gets 
\bea
{\bf g}^{0+}_{L} &=& \f{1}{-\bm{M}_L ~(\omega+i \epsilon)^2+\bm{K}_L^0} \nn \\
{\rm Hence} ~~[{\bf g}^{0+}_{L}]_{l,m} &=& \f{2}{m_L \pi} \int_0^\pi dq \f{ \sin(ql) \sin (qm)}{-(\om +i \epsilon)^2 + \Omega_q^2}~, \nn \\
{\rm where} ~~\Omega_q^2 &=& \f{2 k_L}{m_L} [1- \cos (q) ]  ~.\nn
\eea
We need the $(1,1)^{\rm th}$ element and the above integral gives 
$[{\bf g}^{0+}_{L}]_{1,1}=
{e^{iq}}/{k_L}$, where $q$ is to be obtained from $\om^2=({2 k_L}/{m_L}) (1- \cos q )$.
For the general case $k_0 \neq k_L$, the Green's 
function can be calculated as follows. We write ${\bm K}_L={\bm K}_L^0 + \Delta {\bm K}_L$ where $\Delta {\bm K}_L$ is a 
perturbartion matrix whose the only non-zero element is $\Delta {\bm K}_{11}=k_0 - k_L$. From the definition of the Green's function
$\bm{{\bf g}_L^+}=[-\bm{M}_L (\omega+i \epsilon)^2 +\bm{K}_L^0+\bm{\Delta K}_L]^{-1}$ we get
\bea
\bm{g}^+_{L}+\bm{g}^{0+}_{L}\bm{\Delta K}_L\bm{g}^+_{L}=\bm{g}^{0+}_{L}~.\label{greenpert}
\eea
Taking the $(1,1)^{\rm th}$ element of the above equation gives
\bea
[{\bf g}^+_{L}]_{1,1}=\frac{[{\bf g}^{0+}_{L}]_{11}} {1+[{\bf g}^{0+}_{L}]_{11}~[{\bm {\Delta \bm K}}_L]_{11}} 
=\frac{e^{iq}}{k_L+(k_0-k_L)e^{iq}}~,
\label{g11}
\eea
and similarly
\bea
[{\bf g}^+_{R}]_{1,1}=\frac{e^{iq'}}{k_R+(k_N-k_R)e^{iq'}}~.
\label{g11p}
\eea
Using the definitions given earlier we derive the following expressions:
\bea
\Sigma^+_L &=& \frac{k_0^2 e^{i q}}{k_L+(k_0-k_L) e^{iq}},~~~
 \Sigma^+_R=\frac{k_N^2 e^{i q'}}{k_R+(k_N-k_R) e^{iq'}},~~~ \nn \\
\Gamma_L &=& -\frac{k_0^2 k_L sin(q)}{{|k_0-k_L+k_L ~e^{-iq}|}^2},~~~
 \Gamma_R=- \frac{k_N^2 k_R sin(q^\prime)}{{|k_N-k_R+k_R~e^{-iq^\prime}|}^2}~, \label{gams}
\eea 
where $q,q'$ are respectively obtained from the relations $\om^2=({2 k_L}/{m_L}) (1- \cos q )=({2 k_R}/{m_R}) (1- \cos q')$ and $\Gamma_L,\Gamma_R$ are non-zero only when both $q,q'$ are real.
Plugging in the expressions of $\Sigma^+_L$ and $\Sigma^+_R$ in Eq.~(\ref{det1N-2}), we obtain from Eq.~(\ref{G1N}) 
\bea
{\bm G}^+_{1,N}=\frac{1}{\begin{bmatrix}
                                         1 & -\frac{k_0^2 e ^{iq}}{k_L+(k_0-k_L)e^{iq}}
                                        \end{bmatrix} ~ \hat{T}~
							  \begin{bmatrix}
							    k_N & 0\\
							      0 & 1/k_N\\
							      \end{bmatrix}~ 
							\begin{bmatrix}
							1\\
							\frac{{k_N}^2 e ^{iq^\prime}}{k_R+(k_N-k_R)e^{iq^\prime}}\\
					                                             \end{bmatrix}}~. \label{g1N} 
\eea

\section{Scattering states  and transmission coefficient}
\label{scatt}
For our model the equations of motion correspond to the discrete wave equation 
for which we can construct scattering wave solutions. We will now construct 
solutions that correspond to plane waves incident on the system from the 
reservoirs. From these solutions we will obtain the transmission coefficient. 
In the following we will only consider the ``right-moving states'' which correspond to waves that are incident from the left reservoir. The ``left-moving states'' can be
similarly obtained. 

Let us consider a chain described by the Hamiltonian of Eq.~(\ref{ham1}) with an infinite number of particles in both the reservoirs.
The particle displacements in the chain  satisfy the equations of motion 
\bea
m_l \ddot{x}_l = -(k_{l-1}+k_l)~ x_l +k_{l-1}~x_{l-1} + k_{l}~x_{l+1}~,
\eea
where $l=1,\ldots,N$ refers to particles of the system, $l \leq 0$ refers to particles of the left reservoir ({\emph{i.e}} $\alpha\geq 1$, as
in the notation of Eq. \ref{ham1}), and $l \geq N+1$ refers to particles in the right reservoir  ({\emph{i.e}} $\alpha' \geq 1$). 
We note that these equations are valid both for the quantum representation, 
where the variables are Heisenberg operators, and also for the classical case. Corresponding to the above equations let us 
construct classical wave solutions $\psi_l$ satisfying the  equations    
\bea
m_l \ddot{\psi}_l = -(k_{l-1}+k_l)~ \psi_l +k_{l-1}~\psi_{l-1} + k_{l}~\psi_{l+1}~. \label{waveeq}
\eea
In the left and right reservoirs these equations take  the form of the discrete wave equations
\bea
\ddot{\psi}_\alpha &=& (k_L/m_L) (\psi_{\alpha+1}-2\psi_{\alpha}+\psi_{\alpha-1}) ~~~~~ {\rm for} ~\alpha >1~,
\nn \\
\ddot{\psi}_{\alpha'} &=& (k_R/m_R) (\psi_{\alpha'+1}-2\psi_{\alpha'}+\psi_{\alpha'-1})~~~~ {\rm for}~\alpha' > 1 \nn ~.
\eea
These have the following plane-wave solutions, 
\bea
\psi_\alpha(q)
&= &\f{1}{(2 \pi m_L)^{1/2}} ~e^{-i \omega t}~(~e^{-i q \alpha}+ r ~e^{i q \alpha}~)~~~~~{\rm for} ~\alpha \geq 1~,
\label{wave} \\
\psi_{\alpha'}(q')&=& \f{1}{(2 \pi m_L)^{1/2}} ~\tau ~e^{-i \omega t} e^{i q^\prime \alpha'} ~~~~ {\rm for}~\alpha' \geq 1~,
\eea
where the wave-vectors $q,q'\in (0,\pi)$ satisfy the dispersion relations
\bea
{\omega}^2 =(2 k_L/m_L)~(1-\cos q )= (2 k_R/m_R)(1-\cos q^\prime )~,
\label{dispersion}
\eea
and the normalization is chosen such that for $\tau=0$ ({\emph i.e} no transmission) the following condition is satisfied:   
\bea
\int_0^{\pi} dq~m_L \psi^*_\alpha(q) \psi_\nu(q) = \delta_{\alpha,\nu}~
\eea
for any two points $\alpha,\nu$ on the left bath. 
The solution in Eq.~(\ref{wave}) corresponds to 
a plane wave of wave vector $q$, frequency $\omega$ that is incident on the system from the left side,  part of this is then  
reflected with amplitude $~r$, and a part transmitted across the system with amplitude $\tau$. 
We shall refer to $\tau$ as the transmission coefficient and will now proceed to the calculation of this. As is well-known in quantum mechanics and wave-theory, the required scattering states  can be constructed either 
by    direct solution of the equations of motion in Eq.~(\ref{waveeq}) or by 
the Lippmann-Schwinger scattering theory approach. We now present both these methods. 

\subsection{Transmission coefficient from direct solution of the the wave equation}
For points on the reservoirs the plane wave solution has the form in Eq.~(\ref{wave}). For $ 0 \leq l \leq N+1$, let us write
$\psi_l(q)=s_l e^{-i\omega t}$, $l=0,..,N+1$ where the amplitudes $s_l$ satisfy the equations
\bea
m_l \omega^2 s_l = (k_{l-1}+k_l)~ s_l-k_{l-1}~s_{l-1}-k_{l}~s_{l+1}~,
\eea
and it is to be understood that  $l=0, l=-1$ refer to $\alpha=1,~ \alpha=2$ respectively while  $l=N+1,~l=N+2$ refer to $\alpha'=1,~\alpha'=2$.
Hence we get the following recursion relation:
\bea
\left[ {\begin{array}{cc}
 k_{l-1} s_{l-1}  \\
 s_l \\
 \end{array} } \right] = \hat{T}_l 
\left[ \begin{array}{cc}
        k_{l} s_l\\
        s_{l+1}\\
       \end{array} \right]~,
\eea
where $T_l$ is defined in Eq. (\ref{T_l}).
Using this recursively gives
\bea
\left[ {\begin{array}{cc}
 k_{-1} s_{-1}  \\
 s_0 \\
 \end{array} } \right] = \hat{T}_0 ~\hat{T} ~\hat{T}_{N+1}~ 
\left[ \begin{array}{cc}
        k_{N+1} s_{N+1}\\
        s_{N+2} \\
       \end{array} \right]~. \label{rec1}
\eea
We note that 
\bea
\left[ {\begin{array}{cc}
 k_{-1} s_{-1}  \\
 s_0 \\
 \end{array} } \right]=\f{1}{(2\pi m_L)^{1/2}}
\left[ \begin{array}{cc}
k_L~(e^{-2iq}+r e^{2iq})\\
 (e^{-iq}+r e^{iq}) \\
\end{array} \right]~,~~ 
\left[ \begin{array}{cc}
        k_{N+1} s_{N+1}\\
        s_{N+2} \\
       \end{array} \right]=\f{1}{(2\pi m_L)^{1/2}}
\left[ \begin{array}{cc}
k_R e^{i q'}\\
 e^{2 iq^\prime}
              \end{array} \right] \tau~, \nn \\
\hat{T}_0=\left[ \begin{array}{cc}
k_L+k_0-m_L\omega^2 & -k_0^2 \\
1 & 0\\ 
\end{array} \right],~~~\hat{T}_{N+1}= \left[ \begin{array}{cc}
k_R+k_N-m_R\omega^2 & -k_R^2 \\
1 & 0\\
\end{array}  \right]~, \nn
\eea
and hence Eq.~(\ref{rec1})   gives
\bea
\hat{T}_0~
~ \hat{T}~
\hat{T}_{N+1} ~
\left( \begin{array}{cc}
k_R e^{i q'}\\
 e^{2 iq^\prime}
              \end{array} \right) \tau = 
\left( \begin{array}{cc}
k_L e^{-2iq}\\
e^{-iq} \\
\end{array} \right) + 
\left( \begin{array}{cc}
k_L e^{2iq}\\
 e^{iq} \\
\end{array} \right)~r ~.
\eea
To solve for $\tau$ we multiply the above equation by the row vector $(1~ -k_L e^{iq})$, and this gives
\bea
\tau= \frac{-2 i k_L \sin(q)~ } { (e^{iq}~ -k_L e^{2iq})~ \hat{T}_0
~ \hat{T}~
\hat{T}_{N+1} ~\left( \begin{array}{cc}
k_R e^{i q'}\\
 e^{2 iq^\prime}
              \end{array} \right)}~.\nn
\eea
After 
some simplifications and using the form of ${\bm G}^+_{1,N}$ given in Eq. (\ref{g1N}) we obtain:
\bea
\tau=-\frac{2 i k_L \sin(q)~k_0 k_N e^{-i(q+q^\prime)}}{(k_0-k_L +k_L e^{-iq})(k_N-k_R +k_R e^{-iq^\prime})}
~{\bm G}^+_{1,N} ~.
\label{tau}
\eea
Now using the expressions for $\Gamma_L,\Gamma_R$ in Eq.~(\ref{gams})  and comparing with the formula in Eq.~(\ref{Tw}) we  immediately see that the energy 
transmittance $\mathcal{T}(\omega)$  and the transmission coefficient $\tau(\omega)$ are related as
\bea
\mathcal{T}(\omega)=|\tau(\omega)|^2\frac{k_R \sin(q^\prime)}{k_L \sin(q)}~.
\label{transmittance}
\eea

\subsection{Transmission coefficient from Lippmann-Schwinger scattering approach}

The Lippmann-Schwinger scattering theory approach in quantum mechanics starts by  
breaking up the Hamiltonian of a system into an unperturbed part and a perturbation.  
One then writes an exact scattering solution of the unperturbed part of 
the Hamiltonian  and then uses this to obtain a solution of the full problem 
in terms of the perturbation and appropriate Green's functions. Here, using the 
notation of Eq.~(\ref{ham1D}), we treat $\mathcal{H}_S +\mathcal{H}_L+\mathcal{H}_R$ 
as the unperturbed Hamiltonian and $\mathcal{H}_{LS}+\mathcal{H}_{RS}~$ as the perturbation.    

{\bf Lippmann Schwinger theory}: Let us use the notation $\ket{\psi(q)}$ to denote the state for the wave-function  $\psi_l(q)$ satisfying the wave equation 
\bea
{\bf M} \om^2 \ket{\psi(q)}= {\bf K} \ket{\psi(q)},\label{waveeq2}~,
\eea
where ${\bf M}$ and ${\bf K}$ are the mass matrix and force matrix respectively of the full chain (including system and reservoirs). 
Using the partition of the chain into the reservoir and system parts, these matrices have the following block structures:
\bea
{\bf M}= \begin{bmatrix} 
		  {\bm M}_S  & 0 & 0\\
		  0 &{\bm M}_L &0\\
                  0 & 0 & {\bm M}_R\\
\end{bmatrix},~~~{\bm K}= \begin{bmatrix} 
		  {\bm K}_S & {\bm K}_{SL}  & {\bm K}_{SR}\\
		  {\bm K}_{SL}^T & {\bm K}_L & 0\\
                  {\bm K}_{SR}^T & 0 & {\bm K}_R\\
\end{bmatrix}
\eea
Breaking ${\bm K}$ into unperturbed and perturbed parts we have:
\bea
{\bm K} &=& {\bm K}_0 + {\bm K}_1 \nn \\
{\rm where} &&~~~{\bm K}_0= 
\begin{bmatrix} 
		  {\bm K}_S & 0  & 0\\
		   0 & {\bm K}_L & 0\\
                   0 & 0 & {\bm K}_R\\
\end{bmatrix} ~~,~~~~
{\bm K}_1= \begin{bmatrix} 
		  0 & {\bm K}_{SL}  & {\bm K}_{SR}\\
		  {\bm K}_{SL}^T & 0 & 0\\
                  {\bm K}_{SR}^T & 0 & 0\\
\end{bmatrix}~.
\eea
Treating ${\bm K}_1$ as a perturbation we then obtain the following scattering 
solution of Eq.~(\ref{waveeq2}):
\bea
\ket{\psi} &=& \ket{\psi^0}-{\bm {\mathcal{G}}}^+ {\bm K}_1 \ket{\psi^0}~, \nn \\
{\rm where} ~~{\bm {\mathcal{G}}}^+ &=& [-{\bm M} (\omega +i \epsilon)^2+{\bm K}]^{-1} \label{LSsoln}
\eea 
is the  Green's function for the  full chain and $\ket{\psi^0}$ is a scattering
solution of the unperturbed system satisfying the equation
\bea
{\bf M} \om^2 \ket{\psi^0(q)}= {\bf K}_0 \ket{\psi^0(q)}~.\label{waveequp}
\eea

{\bf Construction of initial state}:  Let us first construct the right-moving scattering states. For this we consider the 
particular initial state $\ket{\psi^0}$ where the left reservoir is in a 
normal mode with frequency $\omega$ while the system and right reservoir degrees of freedom are at rest. 
Thus we choose $\psi^0_l(q)=0$ for $l >0$ and $\psi^0_l(q)=\psi^L_\alpha$ for $l\leq0$, 
with $\alpha=1-l$ and $\psi^L_\alpha(q)$ satisfying the equation
\bea
m_L \om^2 \psi^L_\alpha (q)=\sum_{\beta} [{\bf K}_L]_{\alpha,\beta} {\psi^L_\beta(q)}~.
\eea
The form of ${\bm K}_L$ can be read from  Eq.~(\ref{hamcan}), and we then get
\bea
m_L \om^2 \psi^L_\alpha (q)= k_L [~2 \psi^L_\alpha(q)-\psi^L_{\alpha +1}(q)-\psi^L_{\alpha-1}(q)~]+
\delta_{\alpha,1} (k_0-k_L) \psi^L_1(q)~,~~\alpha=1,2,\ldots~, \label{resmod} 
\eea
with the boundary condition $\psi^L_0(q)=0$. For $k_0=k_L$ the normal modes 
are given by $\psi^L_\alpha(q) = 2i \sin q \alpha/(2 \pi m_L)^{1/2}$, where the normalization is chosen such that 
$\int_0^\pi dq~m_L {\psi^L}^*_\alpha(q) \psi^L_\nu(q)=\delta_{\alpha,\nu}$. For $k_0 \neq k_L$ we can 
 find the normal modes by treating the last term in Eq.~(\ref{resmod}) as 
a perturbation. We will require only $\psi^L_{\alpha=1}(q)$. The 
Lippmann-Schwinger approach is applied again, giving
\bea
\psi^L_{\alpha=1}(q) =\f{2i~\sin(q)}{(2 \pi m_L)^{1/2}}~\big[ 1-(k_0-k_L)~{[{\bf g}^+_L]}_{1,1} \big] =-\f{2i~\sin q}{(2 \pi m_L)^{1/2}} ~k_L e^{-iq}~{[{\bf g}^+_L]}_{1,1},
\label{psi_one}
\eea  
where the result in Eq.~(\ref{g11}) has been used. Note that our choice of $\psi^L_0$ implies an incident wave $e^{i q \alpha}/(2\pi m_L)^{1/2}$.

{\bf Scattering state}: Since we want to finally find $\tau$, it is sufficient to compute the scattering wave function only on the right reservoir. 
From Eq.~(\ref{LSsoln}) we get
\bea
\psi_{\alpha'}(q)=-{\bm {\mathcal{G}}}^+_{\alpha',l=1} k_0 {\psi^L_{\alpha=1}(q)}~. \label{soln_1}
\eea
We now express the Green's function element $-{\bm {\mathcal{G}}}^+_{\alpha',l=1} $ in terms 
of the Green's function ${\bm G}^+$ defined earlier in Eq.~(\ref{GF}).
 We first write  ${\bm {\mathcal{G}}}^+$ in a block-matrix form, with the different 
blocks representing the system and reservoirs. This matrix  satisfies
 the following relation:
\bea 
\begin{bmatrix} 
		  -{\bm M}^S(\omega+i \epsilon)^2+{\bm K}_S & {\bm K}_{SL}  & {\bm K}_{SR}\\
		  {{\bm K}_{SL}}^T & -{\bm M}^L ~(\omega+i \epsilon)^2+ {\bm K}_L & 0\\
                  {{\bm K}_{SR}}^T & 0 & -{\bm M}^R ~(\omega+i\epsilon)^2+ {\bm K}_R\\
\end{bmatrix} \nn \\
~~~~~~~~~~~~\times
\begin{bmatrix}
                 {\bm G}^+_S & {\bm G}^+_{SL} & {\bm G}^+_{SR}\\
		 {\bm G}^+_{LS} & {\bm G}^+_{L} & {\bm G}^+_{LR}\\
		 {\bm G}^+_{RS} & {\bm G}^+_{RL} & {\bm G}^+_{R}\\
\end{bmatrix}
= 
\begin{bmatrix}
               {\bm I} & 0 & 0\\
	       0 & {\bm I} & 0\\
	       0 & 0 & {\bm I}\\
\end{bmatrix}~.
\label{GF_total}
\eea
From this equation the following relations can  be shown to hold \cite{dhar06}: 
\bea
{\bm G}^+_S(\omega)&=&\frac{1}{ -\omega^2{\bm M}^S + {\bm K}_S - {\bm \Sigma}^+_L-{\bm \Sigma}^+_R}=:{\bm G}^+(\omega)~,\\
\label{Gsystem}
{\bm G}^+_{RS}&=&{\bf g}^+_R {\bm K}_{SR} {\bm G}^+~.
\eea
This then gives us
\bea
{\bm {\mathcal{G}}}^+_{\alpha',l=1}= [{\bm G}^+_{RS}]_{\alpha',l=1}
=[{\bf g}^+_R]_{\alpha',1} k_N {\bm G}^+_{N,1}~. \label{reln}
\eea
Using Eqs.~(\ref{psi_one},\ref{reln}) in Eq.~(\ref{soln_1}) we finally get:
\bea
\psi_{\alpha^\prime}(q')=2 i k_0 k_N k_L \sin q e^{-iq} [{\bf g}^+_L]_{1,1} [{\bf g}^+_R]_{\alpha',1}  {\bm G}^+_{1,N}/(2 \pi m_L)^{1/2}~.  
\eea
Now taking the $(\alpha',1)^{\rm th}$ element of Eq.~(\ref{greenpert}), with $L$ replaced by $R$, and using Eq.~(\ref{g11p}) we get
\bea
[{\bf g}^+_R]_{\alpha',1}= \frac{e^{iq' \alpha'}}{k_R+(k_N-k_R)e^{iq'}}~.
\eea 
Using the explicit form of $[{\bf g}^+_{L}]_{1,1}$ from Eq.~(\ref{g11}) we finally 
arrive at the expected form of the transmitted wave function in the right reservoir 
\bea
\psi_{\alpha^\prime}(q^\prime)= \tau e^{iq' \alpha'}/(2 \pi m_L)^{1/2} \nn
\label{transmittedwave}
\eea
with $\tau$ precisely given by the same expression Eq.~(\ref{tau}) obtained in the previous sub-section 
by the direct solution of the wave equation.

\section{Expression for the  energy current in each mode and a derivation of the Landauer formula}
\label{current}
We now use the definition of the heat current operator and show how it can be 
used to express the current contribution of each of the modes in terms of the 
transmission  coefficient and hence the energy transmittance. This will lead 
us to a derivation of the Landauer formula.
In the steady state the current is constant everywhere and we will evaluate 
it  on the right reservoir. Between sites $\alpha'$ and $\alpha'+1$ the 
left-right current is given by the expectation value \cite{dhar08}   
$\hat{j}_L = \langle~ \frac{1}{2}k_R(v_{\alpha^\prime}+v_{\alpha^\prime+1})(x_{\alpha^\prime}-x_{\alpha^\prime+1}) ~\rangle$~ where we compute the average using the $q^{\rm th}$ right-moving state obtained in the previous section. 
It is easiest to obtain this using second-quantized notation. The set of right moving and left moving states form a complete set. Denoting the left-movers by $\psi_{\alpha'}(q)$ with $q \in (-\pi,0)$ we note that they satisfy the completeness relation $\int_{-\pi}^\pi dq (m_{\alpha'})^{1/2} (m_{\nu'})^{1/2} \psi_{\alpha'}^* (q)\psi_{\nu'}(q)=\delta_{\alpha',\nu'}$.  
The displacement and velocity operators at the lattice sites of the right 
bath can be expressed in terms of the creation and annihilation operators $a_{q'},a^\dagger_{q'}$ as
\bea
x_{\alpha^\prime}&=&\displaystyle\int_{-\pi}^\pi d{q} \left(\frac{\hbar }{2 \omega_{q}}\right)^\frac{1}{2}
(a_{q} \psi_{\alpha^\prime}({q})+a_{q}^\dagger 
{\psi_{\alpha^\prime}}^*(q))~, \nn \\
v_{\alpha^\prime}&=& -i \displaystyle\int_{-\pi}^\pi d{q} \left(\frac{\hbar \omega_{q}}{2}\right)^\frac{1}{2}(a_{q}
\psi_{\alpha'}(q)-a_{q}^\dagger {{\psi_{\alpha^\prime}}^*(q)} ) ~.\nn 
\eea
The operators $a_{q_1},a_{q_2}^\dagger$  satisfy the commutation relations 
$[a_{q_1},a^\dagger_{q_2}]=\delta(q_1-q_2)$ and, using the completeness relation, it can be verified that this ensures the usual commutation relations for the position and momentum operators.  
Using the above we get for the expectation value of the current for a right moving state:
\bea
J_{LR}(q) &=& i\hbar k_R (\langle a_{q}^\dagger  a_q \rangle +\frac{1}{2})
[{{\psi_{\alpha^\prime+1}}^*(q)}\psi_{\alpha^\prime}(q)
-\psi_{\alpha^\prime+1}(q){\psi_{\alpha^\prime}}^*(q)]~\nn \\
&=& \frac{\hbar k_R \sin (q')}{\pi m_L}~ |\tau|^2 ~[f(\omega_q,T_L) +\frac{1}{2}]~,
\eea 
where in the last step we have used the form $\psi_{\alpha'}(q)=\tau e^{i \alpha' q'}/(2 \pi m_L)^{1/2}$, and the 
initial occupation probability of the state $q$ is given by the left bath thermal distribution  
 $\langle a_{q}^\dagger a_{q} \rangle= [e^{\hbar \omega_q/k_B T_L}-1]^{-1}=f(\omega,T_L)$.
 The total current transmitted
from the left bath to the right bath, is obtained by integrating over all $q$.
 After making a change of variables from $q$ to $\omega=2(k_L/m_L)^{1/2} \sqrt{1-\cos(q)}$ we get 
\bea
J_{LR} &=& \int_0^\pi dq J_{LR}(q) \nn \\
&&= 
\f{1}{\pi} \int_0^{2(k_L/m_L)^{1/2}} d\omega ~\hbar \omega 
\frac { k_R \sin(q^\prime)}{k_L \sin q}~|\tau|^2 ~ [ f(\omega,T_L)+\frac{1}{2}] \nn \\
&&= \f{1}{\pi} \int_0^{2(k_L/m_L)^{1/2}} d\omega ~\hbar \omega \mathcal{T}(\om) ~ [f(\omega,T_L)+\frac{1}{2}]~,
\eea 
where in the last step we used Eq.~(\ref{transmittance})~.
From symmetry, the current flowing from the right bath to the left bath will be given by
\bea
J_{RL}=\f{1}{\pi} \int_0^{2(k_R/m_R)^{1/2}} d\omega ~\hbar \omega \mathcal{T}(\om) ~ [f(\omega,T_R)+\frac{1}{2}]~.\nn
\eea
Hence finally we get for the net current:
\bea
J=\frac{1}{\pi}\int_{0}^{\om_m} d \omega {\hbar \omega} \mathcal{T}(\omega)
[f(\omega,T_L)-f(\omega,T_R)]~,
\label{jqmech}
\eea 
where $\om_m={\rm min}[2(k_L/m_L)^{1/2},2(k_R/m_R)^{1/2}]$. 
Observing that $\mathcal{T}$ is a symmetric function of $\omega$ and vansishes outside the range $\om \in (0,\om_m)$, we can see 
that Eq.~(\ref{jqmech}) is equivalent to the  Landauer formula Eq.~(\ref{landform})~. 
 
\section{Discussion}
\label{discussion}
In summary we have studied heat conduction across a $1D$ quantum-mechanical 
harmonic chain, with 
arbitrary distribution  of masses and inter-particle spring constants, that is connected to two other ordered $1D$ harmonic crystals which have different mass densities and elastic constants. For this model we use two different approaches to demonstrate the relation $\mathcal{T}(\omega)=(k_R \sin q^\prime)~ |\tau(\omega)|^2 /(k_L \sin q )$ between the
energy transmittance $\mathcal{T}(\omega)$, which occurs in the 
Landauer formula for heat current,  and the transmission 
coefficient $\tau(\omega)$ related to passage of plane waves across the system. 
In the first approach we use the fact that the Green's function occuring in the 
expression for $\mathcal{T}$  has a simple representation in terms of product of $2\times 2 $ matrices. The plane wave solutions are then obtained by directly 
solving the equations of motion and a representation of $\tau$ is 
obtained,  again  in terms of the product of matrices. The connection between $\mathcal{T}$ and $\tau$ is then directly obtained.  
This approach can be extended to the case of regular lattices using, for example,  the techniques used in \cite{chaudhuri10} for the representation of the Green's functions using matrix products. 

In the second approach it is not necessary to find the explicit form of the 
Green's function. One notes that the required plane wave scattering states can be obtained by using the Lippmann-Schwinger approach to evolve initial states  
which are eigenmodes of either one of the reservoirs and are  initially 
localized within the reservoirs. The Lippmann-Schwinger approach then directly 
gives $\tau$ in terms of the Green's function. 
This second approach is more powerful since it can be used for arbitrary harmonic structures where it is not possible to think of simple plane wave scattering 
states. This approach tells us that we need to construct scattering states by 
evolving  the eigenmodes of the two isolated reservoirs. Indeed this is what 
the NEGF approach does in effect and our explicit calculations for a simple but representative model clarifies the picture.  Our exact calcultions also 
illustrate some of the subtle points involved, such as the correct computation 
of the self-energies $\Sigma^+_L,\Sigma^+_R$ for inhomogeneous chains, 
appropriate normalizations of normal-modes  and the choice of initial states.

\end{document}